\documentclass[aps,prl,twocolumn,showpacs,superscriptaddress]{revtex4}
\usepackage{graphicx}

\begin{document}

\title{Controlling Energy Gap of Bilayer Graphene by Strain}
\author {Seon-Myeong Choi}
\affiliation{Department of Physics, Pohang University of 
Science and Technology, Pohang 790-784, Korea}
\author {Seung-Hoon Jhi}
\email{jhish@postech.ac.kr}
\affiliation{Department of Physics, Pohang University of 
Science and Technology, Pohang 790-784, Korea}
\affiliation{Division of Advanced Materials, 
Pohang University of Science and Technology, Pohang 790-784, Korea}
\author {Young-Woo Son}
\email{hand@kias.re.kr}
\affiliation{Korea Institute for Advanced Study, Seoul 130-722, Korea}
\date{\today}
\begin{abstract}
Using the first principles calculations, we show that mechanically tunable electronic energy gap 
is realizable in bilayer graphene if different homogeneous strains are applied to the two layers. 
It is shown that the size of energy gap can be simply controlled by adjusting the strength 
and direction of these strains. 
We also show that the effect originates from the occurrence 
of strain-induced pseudo-scalar potentials in graphene. 
When homogeneous strains with different strengths 
are applied to each layer of bilayer graphene, 
transverse electric fields across the two layers can be generated 
without any external electronic sources, thereby opening an energy gap. 
The results demonstrate a simple mechanical method of realizing pseudo-electromagnetism 
in graphene and suggest a maneuverable approach to 
fabrication of electromechanical devices based on bilayer graphene.
\end{abstract}
\maketitle
Materials that can change energy gaps under mechanical deformations are greatly desired to develop
electromechanical nanodevices with controllable optical and electrical operational ranges~\cite{crespi97,tombler00}.
Graphene, one atom thick semi-metallic membrane~\cite{novoselov05,zhang05,castro09},
may have this property because of its special
electromechanical characteristics such as strain-induced phases in 
the chiral wavefunctions of the massless Dirac 
Hamiltonian~\cite{castro09,fogler08,pereira09l,pereira09b,guinea09,guinea10}.
Although external electric fields have been used to tune the energy gaps in several forms of
graphene~\cite{mccann06,ohta06,son06,min07,castro07,oostinga08,kuzmenko09,maik09,li09,zhang09}, 
a simple and practical method of tuning the electronic
energy gaps based on mechanical operations is lacking.

The influence of smooth elastic deformations on the physics of a single layer of graphene can be
described by introducing a suitable gauge-field vector potential to the free massless particle
Dirac's equation~\cite{castro09}. The choice of vector potential depends on the direction and
strength of applied strains. Specifically designed inhomogeneous strains can produce a net pseudo-magnetic
flux that be expected to induce interesting electromagnetic
effects~\cite{fogler08,pereira09l,pereira09b,guinea09,guinea10}. In contrast, under homogeneous
uniaxial or biaxial strains, the effect of pseudo-magnetic fields is absent but a homogeneous
pseudo-scalar potential can be generated owing to electron density
variations or dilations~\cite{guinea10,ono66,suzuura02,choi10}. Because graphene is a one atom thick semimetal,
the pseudo-scalar potential will alter its work function significantly. When a uniaxial (or isotropic)
strain is applied up to an experimentally realizable magnitude of $\sim$10\%, the work function of graphene
is predicted to increase by 0.27 (or 0.65) eV from the value of 4.49 eV of graphene in equilibrium~\cite{choi10}.
Hence, the homogeneous strain can have important applications in various electronic devices for tuning
the band lineup between metallic leads and graphene, or in changing the amount of charge transfer
from foreign molecules. We also expect that such strain-induced pseudo-scalar potentials will
modify the electronic structure of bilayer graphene significantly.

In this paper, we predict the formation of mechanically 
tunable electronic energy gap in bilayer graphene.
It is shown that the perpendicular electric fields across
the two layers of bilayer graphene can be generated without
any external gate potential if each layer is subjected to 
different strengths of homogeneous strains.
This effect originates from asymmetric generations
of pseudo-scalar potential in each layer of strained bilayer graphene, 
which alter their work functions significantly~\cite{son06}.
We also show that the strain-induced electronic energy
gaps are indirect in general and strongly depend on 
the size and direction of applied homogeneous strains.
We discuss origins of such peculiar variations of the bandgaps and
present realistic experimental setups for observing the energy gaps.

Our study of electronic structures of strained bilayer graphene 
is based on the pseudopotential density functional method~\cite{kresse94}. 
The exchange-correlation interactions were treated within the local density approximation~\cite{lda}.
The cutoff energy for expansion of wavefunctions and potentials was 400 eV.
Extensive momentum space sampling in the whole
Brillouin zone (BZ) is performed to obtain accurate energy dispersions, reliable energy gaps,
and their dependences on strain. 
The Monkhorst-Pack $k$-point grid of $2\times12\times1$ is used for the atomic relaxation and of $5\times30\times1$
for electronic structure calculation of 11.1\% asymmetrically strain bilayer graphene.
Calculations for other strengths of strain also used an equivalent number of $k$-points.
The atomic relaxation was carried out until the Helmann-Feynman forces were less than 0.03 eV/\AA.
The distance between adjacent bilayer graphene in the supercells is 26\AA~avoiding spurious
interactions between them.
We also have increased the distance up to 36{\AA} finding no changes in electronic structures.
The weak dispersion force between the two layers of bilayer graphene is not considered here and the interlayer
distance is fixed to 3.35{\AA} because its effect is negligible when obtaining the energy bands using
the present calculation scheme~\cite{barone08}. We note that present methods correctly describe the trend of
the energy gap in bilayer graphene under external electric fields with a typical underestimate of
the gap size compared with a recent optical measurement~\cite{zhang09}.

\begin{figure}[t]
\includegraphics[width=1.00\columnwidth]{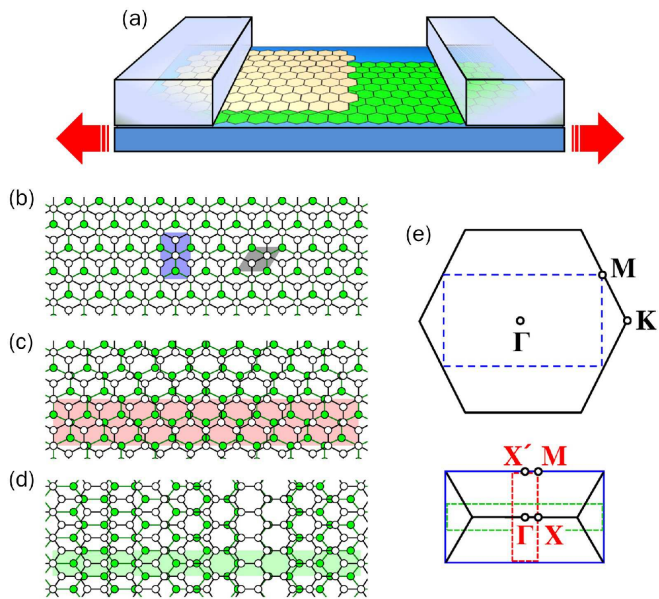}
\caption{(Color online)
(a) Bilayer and single layer graphene junction between the source and drain. 
Red arrows: tensile force direction. 
(b) Schematic atomic structure of bilayer graphene at equilibrium, 
showing carbon atoms in the top layer (clear circles) and the bottom
layer (green circles), and the bilayer graphene unit cells with four carbon atoms (grey parallelogram)
and  eight carbon atoms (blue rectangle). 
(c) Atomic structure of asymmetrically strained bilayer along the 
zigzag chain direction (AZS) with 11.1\% strain on the bottom layer. 
(d) Asymmetrically strained bilayer in armchair direction (AAS) 
with 20\% strain on the bottom layer. 
(e) Top panel: the first Brillouin zones (BZ) of bilayer 
graphene in equilibrium for the unit cell with four carbon
atoms (black hexagon) and for the unit cell with eight carbon atoms (blue dotted rectangle).
Bottom panel: first BZ  for the AZS in (c) (red dotted rectangle); first BZ for AAS bilayer in (d)
(green dotted rectangle).}
\label{fig1}
\end{figure}

Homogeneous strains can be applied to the two layers of bilayer graphene asymmetrically
in various device configurations. The prototypical system considered here is a junction between
single-layer graphene and bilayer graphene [Fig.~\ref{fig1}(a)]; this junction can be synthesized
using the graphene transfer technique~\cite{kim09}, and many mechanically exfoliated samples
contain such junction geometries naturally~\cite{blake07}. If a uniform tensile force is applied
to the junction [Fig.~\ref{fig1}(a)], each layer in bilayer graphene experiences a different magnitude
of homogeneous strain, because only one end of the top layer in bilayer graphene is pinched
by the metallic lead whereas the other end is free-standing [Fig.~\ref{fig1}(a)]. In this case, the top layer
of bilayer graphene in the junction maintains its equilibrium atomic structure due to the negligible
interlayer interaction along the in-plane direction~\cite{kolmogorov04}, but the bottom layer is
subjected to biaxial strain because both of its ends are clamped to the stretched substrates
by the metallic leads [Fig.~\ref{fig1}(a)]. Generally speaking, any device structure with asymmetric
contacts to the top and bottom layers can produce asymmetric strains in each layer due to external
tensile forces. Hereafter, we apply the periodic boundary condition to the asymmetrically strained
bilayer graphene only. To satisfy the commensurability requirement for between the top and bottom
layers along the direction of applied tensile force, the unit cell of the system must be a rectangle
with a very high aspect ratio [Fig.~\ref{fig1}(c) and ~\ref{fig1}(d)]. Accordingly, the first BZ of the
strained graphene is folded into a small rectangle within the first BZ of the original bilayer
graphene in equilibrium [Fig.~\ref{fig1}(e)].

Here we considered two atomic configurations for asymmetrically strained bilayer graphene.
Because pristine bilayer graphene has Bernal stacking between top and bottom layers [Fig.~\ref{fig1}(b)],
we apply strains along the two highly symmetric (i.e., zigzag and armchair) crystallographic
directions of the hexagonal lattice of the bottom layer. First we fix the stacking order along the
short edge of the unit cell, and then stretch the bottom layer along the normal to the edge.
When the bottom layer is under strain along the zigzag-chain direction, we call this configuration
the asymmetrically Z-strained bilayer graphene (AZS bilayer) [Fig.~\ref{fig1}(c)]. When strain is
applied along the armchair-chain direction in the bottom layer, we call it the asymmetrically
A-strained bilayer graphene (AAS bilayer) [Fig.~\ref{fig1}(d)]. We note that, in single-layer graphene,
strains along those two directions induce different variations in its electronic structures
and the effect of strain along arbitrary directions can be represented as a linear combination
of them~\cite{pereira09b,choi10}.

\begin{figure}[t]
\includegraphics[width=1.00\columnwidth]{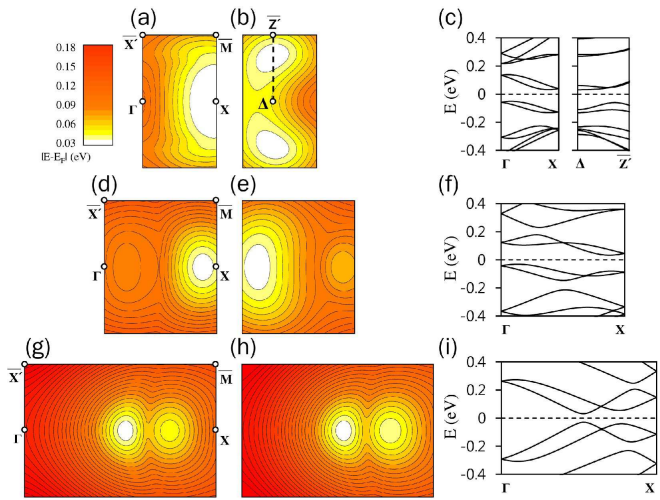}
\caption{(Color online) The energy contours for 
(a) the conduction band and (b) the valence band of the AZS bilayer
with 4\% strain on the bottom layer. Only contours for a part of the first BZ are drawn, where
energy bands exhibit minimal and maximal values. 
Here, $\Gamma\bar{\textrm{X}}'$ is one tenth of $\Gamma{\textrm{X}'}$ in the first BZ shown 
in Fig.~\ref{fig1}(f). 
The scale bar for the contour is given with respect to the Fermi energy ($E_F$) 
in units of eV. 
(c) the electronic band structure of the AZS bilayer with 4\% strain (left panel)
along the high symmetric line ($\Gamma\textrm{X}$) and (right panel) along the dotted line in (b) that crosses
the k-points at which the minimum indirect band gap occurs. 
(d), (e) and (f) Energy contours and band
structures for the AZS bilayer with 6.25\% strain using the same convention as in (a), (b) and (c).
(g), (h) and (i) Energy contours and bands for the AZS bilayer with 11.1\% strain with the same
previous convention.}
\label{fig2}
\end{figure}

In AZS bilayer graphene, an energy gap opens immediately when
the bottom layer is stretched along the zigzag chain direction; the position where the energy gap
opens in the first BZ shifts depends on the magnitude of strain and is not located at the highly
symmetric points of the first BZ [Fig.~\ref{fig2}]. The fundamental energy gap is indirect at
strains up to 9\%, but direct at greater strains. In AAS bilayer graphene, no energy gap opening
occurs up to a strain of about 14\% (not shown here). 
This shows that electronic band gap of asymmetrically
strained bilayer graphene strongly depends on the direction of strain.

\begin{figure}[t]
\includegraphics[width=1.00\columnwidth]{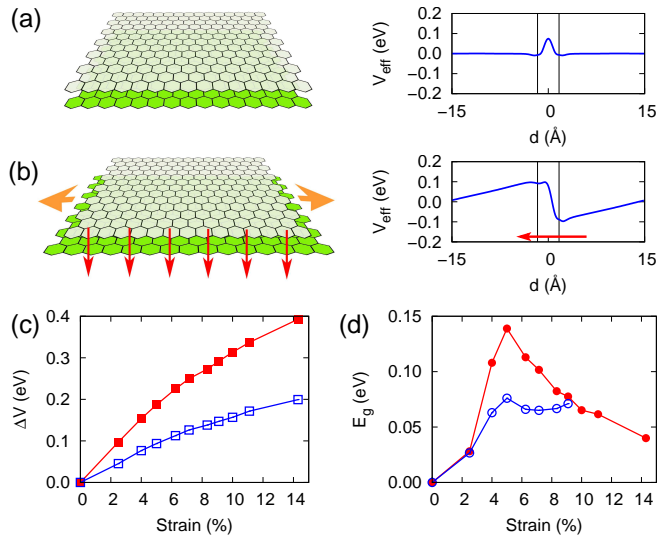}
\caption{(Color online) (a) Bilayer graphene in equilibrium (left panel)
and the effective potential for electrons (right panel). Two vertical lines denote the position of bottom
and top graphene layers, respectively. (b) Asymmetrically strained bilayer graphene with the bottom
layer uniformly stretched along the orange arrows. 
A uniform electric field across the two layers (red arrows) 
is generated due to work function difference (left panel). The effective potential for electrons for AZS
bilayer with 14.3\% strain on the bottom layer (right) and corresponding
electric field (red arrow) across the two layers  are shown. (c) The work function difference (red)
and the net potential difference (blue) between the top and bottom layers as a function of strain (\%)
for the AZS bilayer. (d) The direct (red) and indirect (blue) energy gap of the AZS bilayer
as a function of strain.}
\label{fig3}
\end{figure}

The energy gap of AZS bilayer graphene is created by strain-induced perpendicular electric fields
that break the symmetry in the onsite energy of the top and bottom layers~\cite{mccann06,ohta06}.
The transverse electric fields across the two layers without any external electric source are
possible because the increase of the work function of the strained bottom graphene generates a net
charge transfer from the top layer to the bottom layer [Fig.~\ref{fig3}]. The electric fields
shift and hybridize the conduction and valence bands, thereby opening energy gaps at the points
at which band mixing occurs at the Fermi level. The gap opening in this system is thus a truly
physical manifestation of the pseudo-scalar potential induced by mechanical deformations of graphene.
In our calculations, the work function of single layer graphene increases almost linearly (0.4 eV by 14\%
strain) and accordingly, so does the net charge transfer between the two layers [Fig.~\ref{fig3}(c)].
Due to the screening of the transferred charge, the net difference between local potentials in the
two layers in bilayer graphene increases by 0.2 eV for when asymmetric strain is 14\% [Fig.~\ref{fig3}(c)].
To see the change of onsite energies of each layer clearly, we calculate 
the effective potentials for electrons in the systems by substracting the total
potential of each single-layer graphene from that of bilayer graphene and then
taking average in the plane [right panels in Figs.~\ref{fig3} (a) and (b)].
The induced dipole field across the two layers is clearly visible when the lower layer
is under the strain of 14.3\% as shown in the right panel of Fig.~\ref{fig3} (b).

The indirectness of the energy gap in the AZS bilayer originates from the inequivalent formation
of strain-induced vector potentials in the two graphene layers. In single layer graphene,
homogeneous strains shift the Dirac cones away from the high symmetric points in the first BZ,
and this effect is equivalent to introducing a constant vector potential in the Dirac
Hamiltonian~\cite{castro09,pereira09b,choi10}. The constant vector potential usually has no
effect because it can be simply gauged away. In the present case, only the bottom layer has
such a constant vector potential and it cannot be trivial. The displacement of Dirac cones in the
strained bottom layer alters the points of band crossing between the conduction and valence bands
of the bilayer, which should meet at the K-points of the first BZ of bilayer graphene without
such asymmetric strains~\cite{castro09,mccann06}. Our calculations, suggest that asymmetrically
strained bilayer graphene is unique in that both pseudo-scalar and pseudo-vector potentials induced
by the strain open the energy gap and determine its characteristics.

The competition between various channels of interlayer interactions is yet another important factor
that determines the size of the energy gap. All AAS bilayers and the highly strained AZS bilayer
contain large domains of carbon hexagons that have almost a perfect overlap (AA stacking~\cite{horiuchi03})
between pairs of layers [Fig.~\ref{fig1}(d)]. In bilayer graphene with AA stacking, the low energy
electrons behave as those in two interpenetrating Dirac cones of single layer graphene and no energy gap
forms even with strong transverse external electric fields~\cite{nanda09}. Likewise, when the interlayer
interactions in the AA stacking dominate over all other interactions, the significant differences
in the work function between the top and bottom layers are less effective in opening energy gaps
than the cases with the Bernal stacking. For this reason, when the strain exceeds 9\% in the AZS
bilayer, the energy gap starts to decrease even though the net potential difference continues to
increase. Moreover, when strain is applied along the armchair direction in the bottom layer of
the AAS bilayer, the interlayer interaction in the AA stacking domain is predominant over all other
channels and the energy gap does not open at all [Fig.~\ref{fig1}(d)].

We find that as the strain in the bottom layer increases, the energy gap of the AZS bilayer first
increases linearly, then decreases with a transition from an indirect gap to a direct gap. At a strain of 5\%,
both the indirect energy gap (0.076 eV) and the direct energy gap (0.14 eV) are maximal [Fig.~\ref{fig3}(d)].
This variation of energy gaps depending on the magnitude of asymmetric strains in bilayer graphene is
quite in contrast to that observed in simple Bernal-type stacked bilayer graphene under external
electric fields~\cite{mccann06,ohta06,min07,zhang09}. The latter shows a monotonic increase and then
saturation at an energy gap of $\sim$ 0.25 eV due to screening of induced charges~\cite{mccann06,min07,zhang09}
as external fields increase. The unique interplay between atomic stacking and strain-induced
pseudo-scalar potential leads to the maximal energy gap at a certain magnitude and direction of
strain. Hence, even under very gentle strains, the rectifying current due to the strain-induced
energy gap should be measured in the single-layer/bilayer graphene junctions considered here.
The atomic orientation of bilayer graphene and the size of applied strains could also be inferred
from such measurements.

In summary, we show that one can control the energy gap of bilayer graphene 
only by the strength and direction of homogeneous strain within 
the experimentally accessible range~\cite{kim09,hong10}.
We show that, if homogeneous strains with different strengths are 
applied to each layer of bilayer graphene, transverse electric fields 
across the two layers can be generated without any external electronic sources, 
thereby opening an energy gap and enabling the system to behave 
as an electromechanical switch with tunable energy band gaps. 
This finding is a materialization of pseudo-vector and pseudo-scalar potentials 
induced by external mechanical perturbations
in the relativistic massless Dirac Hamiltonians. 

We thank H. J. Lee, B. H. Hong, and H.-Y. Kee for discussions and comments and specially
thank S. Y. Jung for discussions on device geometries. S.-M. C. and S.-H. J. were supported
by the National Research Foundation of Korea (NRF) grant funded by the Korea government (MEST)
(Grant No. 2009-0087731 and WCU program No. R31-2008-000-10059-0). Y.-W. S. was supported
by the NRF grant funded by MEST (Quantum Metamaterials research center,
No. R11-2008-053-01002-0 and Nano R\&D program 2008-03670).


\begin{thebibliography}{99}
\bibitem{crespi97} V. H. Crespi, M. L. Cohen, and A. Rubio, Phys. Rev. Lett. {\bf 79}, 2093 (1997).
\bibitem{tombler00} T. W. Tombler {\it et. al.}, Nature {\bf 405}, 769 (2000).
\bibitem{novoselov05} K. S. Novoselov {\it et. al.}, Nature {\bf 438}, 197 (2005).
\bibitem{zhang05} Y. Zhang, Y.-W. Tan, H. L. Stormer and P. Kim, Nature {\bf 438}, 201 (2005).
\bibitem{castro09} A. H. Castro Neto, F. Guinea, N. M. R. Peres, K. S. Novoselov and A. K. Geim, Rev. Mod. Phys. {\bf 81}, 109 (2009).
\bibitem{fogler08} M. M. Fogler, F. Guinea and M. I. Katsnelson, Phys. Rev. Lett. {\bf 101}, 226804 (2008).
\bibitem{pereira09l} V. M. Pereira, V. M. and A. H. Castro Neto, Phys. Rev. Lett. {\bf 103}, 046801 (2009).
\bibitem{pereira09b} V. M. Pereira, A. H. Castro Neto and N. M. R. Peres, Phys. Rev. B {\bf 80}, 045401 (2009).
\bibitem{guinea09} F. Guinea, M. I. Katsnelson and A. K. Geim, Nature Phys. {\bf 6}, 30 (2009).
\bibitem{guinea10} F. Guinea, A. K. Geim, M. I. Katsnelson and K. S. Novoselov, Phys. Rev. B {\bf 81}, 035408 (2010).
\bibitem{mccann06} E. McCann, Phys. Rev. B {\bf 74}, 161403(R) (2006).
\bibitem{ohta06} T. Ohta {\it et. al.}, Science {\bf 313}, 951 (2006).
\bibitem{son06} Y.-W. Son, M. L. Cohen and S. G. Louie, Nature {\bf 444}, 347 (2006).
\bibitem{min07} H. Min, B. Sahu, S. K. Banerjee and A. H. MacDonald, Phys. Rev. B {\bf 75}, 155115 (2007).
\bibitem{castro07} E. V. Castro {\it et. al.}, Phys. Rev. Lett. {\bf 99}, 216802 (2007).
\bibitem{oostinga08} J. B. Oostinga {\it et. al.}, Nature Mat. {\bf 7}, 151 (2008).
\bibitem{kuzmenko09} A. B. Kuzmenko {\it et. al.}, Phys. Rev. Lett. {\bf 103}, 116804 (2009).
\bibitem{maik09} K. F. Maik, C. H. Lui, J. Shan, and T. F. Heinz, Phys. Rev. Lett. {\bf 102}, 256405 (2009).
\bibitem{li09} Z. Q. Li {\it et. al.}, Phys. Rev. Lett. {\bf 102}, 037403 (2009).
\bibitem{zhang09} Y. Zhang {\it et. al.}, Nature {\bf 459}, 820 (2009).
\bibitem{ono66} S. Ono and K. Sugihara, J. Phys. Soc. Jpn. {\bf 21}, 861 (1966).
\bibitem{suzuura02} H. Suzuura and T. Ando, Phys. Rev. B {\bf 65}, 235412 (2002).
\bibitem{choi10} S.-M. Choi, S.-H. Jhi and Y.-W. Son, Phys. Rev. B, {\bf 81}, 081407(R) (2010).
\bibitem{kresse94} G. Kresse and J. Hafner, Phys. Rev. B {\bf 49}, 14251 (1994).
\bibitem{lda}D. M. Ceperley and B. J. Alder, Phys. Rev. Lett. {\bf 45}, 566 (1980).
\bibitem{barone08} V. Barone {\it et. al.}, J. Comput. Chem. {\bf 30}, 934 (2008).
\bibitem{kim09} K. S. Kim {\it et. al.}, Nature {\bf 457}, 706 (2009).
\bibitem{blake07} P. Blake {\it et. al.}, Appl. Phys. Lett. {\bf 91}, 063124 (2007).
\bibitem{kolmogorov04} A. N. Kolmogorov and V. H. Crespi, Phys. Rev. Lett. {\bf 85}, 065503 (2004).
\bibitem{horiuchi03} S. Horiuchi {\it et. al.}, Jpn. J. Appl. Phys. {\bf 42}, L1073 (2003).
\bibitem{nanda09} B. R. K. Nanda and S. Satpathy, Phys. Rev. B {\bf 80}, 165430 (2009).
\bibitem{hong10} S. Bae {\it et. al.} Nat. Nanotechnol. {\bf 5}, 574 (2010).
\end{thebibliography}
\end{document}